\documentstyle{article}

\begin{document}

\begin{center}
{\large {\bf Resonant Absorption between Moving Atoms}}

{\large {\bf due to Doppler Frequency Shift and Quantum Energy Variation}}

$\ $

{\sf Ching-Chuan Su}

Department of Electrical Engineering

National Tsinghua University

Hsinchu, Taiwan
\end{center}

$\ $

\noindent {\bf Abstract }-- By taking both the Doppler frequency shift for
electromagnetic wave and the quantum energy variation of matter wave into
consideration, a resonant-absorption condition based on the local-ether wave
equation is presented to account for a variety of phenomena consistently,
including the Ives-Stilwell experiment, the output frequency from ammonia
masers, and the M\"{o}ssbauer rotor experiment. It is found that in the
resonant-absorption condition, the major term associated with the laboratory
velocity is a dot-product term between this velocity and that of the
emitting or absorbing atom. This term appears both in the Doppler frequency
shift and the transition frequency variation and then cancels out. Thereby,
the experimental results can be independent of the laboratory velocity and
hence comply with Galilean relativity, despite the restriction that the
involved velocities are referred specifically to the local-ether frame.
However, by examining the resonant-absorption condition in the M\"{o}ssbauer
rotor experiment to a higher order, it is found that Galilean relativity
breaks down.

$\ $

$\ $

\noindent {\large {\bf 1. Introduction}}

It is well known that the Doppler effect due to the relative motion between
a transmitter and a receiver causes a shift in the received frequency of
electromagnetic or acoustic wave. The Doppler effect has been applied by
Ives and Stilwell to deal with the frequency shift in the light emitted from
a fast-moving hydrogen atomic beam [1]. However, it was found that the
observed frequency shift agrees with the Doppler effect only to the first
order of the atom speed normalized to the speed of light $c$. In order to
account for the observed shift correctly to the second order, a hypothesis
of Larmor and Lorentz was adopted, which states that the frequency of a wave
radiated from a moving source of speed $v$ is altered by a factor of $\sqrt{%
1-v^{2}/c^{2}}$ [1]. It is noted that this speed-dependent factor is
identical to the one in the famous Lorentz mass-variation law. Presently, an
almost unanimously accepted explanation of this additional frequency shift
is provided by Einstein's special relativity, whereby a second-order Doppler
effect has been derived based on the Lorentz transformation of space and
time [2]. Thus it seems to imply that the frequency shift to the second
order between a transmitter and a receiver in relative motion is purely a
kinematical property. And, for the case of resonant absorption between
moving atoms or ions, it seems to assume tacitly that the transition
frequencies of the emitting or the absorbing atoms in motion remain
unchanged, in spite of the well-known fact that the electronic quantum
states in atoms depend on the mass of electron, which in turn depends on the
particle speed. Furthermore, in the common understanding the speed $v$ which
determines the second-order Doppler frequency shift by the mass-variation
factor is the relative speed between the transmitter and the receiver.
However, in the Hafele-Keating experiment with circumnavigating atomic
clocks and in the GPS (global positioning system) with atomic clocks onboard
the orbiting satellites, it has been demonstrated with a high precision that
the speed which determines the transition frequency and clock rate by the
mass-variation factor is referred uniquely to an ECI (earth-centered
inertial) frame [3]. Thus there seems to exist a discrepancy in the
reference frame for the speed in the mass-variation factor.

Recently, we have presented the local-ether model of propagation of
electromagnetic wave [4]. It is supposed that electromagnetic wave
propagates via a medium like the ether. However, the ether is not universal.
Specifically, it is supposed that in the region under sufficient influence
of the gravity due to the Earth, the Sun, or another celestial body, there
forms a local ether which as well as the associated gravitational potential
moves with the respective body. Within each local ether, it is proposed that
as in the classical propagation model, electromagnetic wave propagates at
the speed of light $c$ with respect to the associated local ether,
independent of the motions of source and receiver. Thereby, for earthbound
or interplanetary waves, the propagation is referred specifically to a
geocentric or a heliocentric inertial frame, respectively. This local-ether
model has been adopted to account for the effects of earth's motions in a
wide variety of propagation phenomena, particularly the Sagnac correction in
GPS, the time comparison via intercontinental microwave link, and the echo
time in interplanetary radar. As examined within the present accuracy, the
local-ether model is still in accord with the Michelson-Morley experiment
which is known to make the classical ether notion obsolete. Furthermore, by
modifying the speed of light in a gravitational potential, this simple
propagation model leads to the deflection of light by the Sun and the
increment in the interplanetary radar echo time which are important
phenomena supporting the general theory of relativity. Moreover, based on
this new classical model, the first-order Doppler frequency shift has been
derived to account for the anisotropy in antenna temperature of CMBR (cosmic
microwave background radiation), the shift in the spectrum of light radiated
from a moving star, and the seasonal variation of eclipse intervals in
Roemer's observations of Jupiter's moons [4].

Further, the matter wave associated with a particle has been supposed to
follow the local-ether model and be governed by a wave equation
incorporating a natural frequency and the electric scalar potential [3].
From the local-ether wave equation, a first-order time evolution equation
similar to Schr\"{o}dinger's equation is derived. From the electrostatic
force derived from this evolution equation, it has been found that the rest
mass of a particle is just the natural frequency, aside from a scaling
factor. That is, the inertial mass of a particle originates from the natural
frequency and hence from the temporal variation of the associated matter
wave. Furthermore, due to the dispersion of matter wave, it has been derived
that the mass of the particle and the angular frequency of the matter wave
increase with the particle speed by the famous Lorentz mass-variation
factor, except that the speed is referred specifically to the associated
local-ether frame. A feature different from Schr\"{o}dinger's equation is
that the time derivative in the local-ether evolution equation incorporates
an extra multiplying term of the ratio of the speed-dependent angular
frequency to the natural frequency. As a consequence, it has been found that
the energies of quantum states of the matter wave bounded in an atom
decrease with the inverse of the speed-dependent mass [3]. Thus the
transition frequency and the atomic clock rate decrease with the atom speed
by the mass-variation factor. Thereby, the frequency variation associated
with the speed-dependent mass-variation factor has been derived as an
intrinsic quantum property of the matter wave bounded in atom. Similarly,
the gravitational redshift has been derived as an effect of the
gravitational potential on the quantum energy. Anyway, the dependence of
quantum energy on speed can be expected at least for hydrogen-like atoms.
Since it is well known in standard textbooks on quantum mechanics that the
electronic quantum states in such an atom depend on the mass of electron,
which in turn has been known to depend on speed. However, according to the
local-ether model, the atom speed that determines the mass variation is
referred uniquely to an ECI (earth-centered inertial) frame for earthbound
atoms. Thus the quantum energy and hence the atomic clock rate tend to
depend on earth's rotation, but is entirely independent of earth's orbital
motion around the Sun or others. This consequence has been used to account
for the east-west directional anisotropy in the Hafele-Keating experiment,
the synchronism among GPS atomic clocks, and for the spatial isotropy in the
Hughes-Drever experiment [3].

In this investigation, based on the local-ether propagation model, a
higher-order Doppler frequency shift for electromagnetic wave is derived.
Further, by taking the speed-dependent variation of quantum energy of matter
wave into account, we present a resonant-absorption condition between moving
emitting and absorbing atoms. Then this frequency relation is applied to
deal with various experiments consistently, including the Ives-Stilwell
experiment, the output frequency from ammonia masers, and the M\"{o}ssbauer
rotor experiment. Moreover, we examine the spatial isotropy, Galilean
relativity, and their breakdowns in these experiments.

$\ $

\noindent {\large {\bf 2. Higher-Order Doppler Frequency Shift}}

In this section, based on the local-ether model of wave propagation, the
Doppler frequency shift is given to higher orders. Consider the case where
both the source and the receiver are located within the same local ether and
are moving at velocities ${\bf v}_{s}$ and ${\bf v}_{e}$ with respect to
this local-ether frame, respectively. According to the classical propagation
model, the propagation time $\tau $ is the propagation range $R$ divided by
the isotropic speed $c$ as $\tau =R/c$. It is important to note that the
propagation range is the distance from the position of the source at the
instant of wave emission to the one of the receiver at the instant of
reception, as viewed in the local-ether frame. Thus the propagation range
and hence the propagation time depend on the velocity ${\bf v}_{e}$ and the
acceleration ${\bf a}_{e}$ of the receiver with respect to the local-ether
frame. Quantitatively, when $\tau $ is short, the propagation range is given
implicitly by 
$$
R=\left| {\bf R}_{t}+{\bf v}_{e}\frac{R}{c}+{\bf a}_{e}\frac{R^{2}}{2c^{2}}%
\right| ,\eqno
(1) 
$$
where ${\bf R}_{t}$ is the directed propagation-path length from the source
to the receiver both at the instant of emission. It can be shown that to the
second order of normalized speed, the propagation range can be given
explicitly in terms of the path length $R_{t}$ by [4] 
$$
R=R_{t}\left\{ 1+\frac{1}{c}u_{e}+\frac{1}{2c^{2}}\left( u_{e}^{2}+v_{e}^{2}+%
{\bf a}_{e}\cdot {\bf R}_{t}\right) \right\} ,\eqno
(2) 
$$
where radial speed $u$ ($={\bf v}\cdot \hat{R}_{t}$) is the component of
velocity ${\bf v}$ along ${\bf R}_{t}$ and the unit vector $\hat{R}_{t}={\bf %
R}_{t}/R_{t}$.

Consider the Doppler effect for the case where the source is emitting wave
periodically and is moving with respect to the receiver. Due to the motions
of the transmitter and receiver, the rate of reception tends to be different
from the one of emission. As discussed in [4], the received time difference $%
\triangle t$ between two signals transmitted with a differential time
difference $\triangle t^{\prime }$ can be given in terms of the difference
in the propagation range by 
$$
\triangle t=\triangle t^{\prime }+\frac{R(t^{\prime }+\triangle t^{\prime })%
}{c}-\frac{R(t^{\prime })}{c},\eqno
(3) 
$$
where $R(t)$ denotes the propagation range for the wave emitted at an
arbitrary instant $t$. Then the received frequency $f_{r}$ and the
transmitted frequency $f_{t}$ are related by 
$$
\frac{f_{t}}{f_{r}}=\frac{\triangle t}{\triangle t^{\prime }}=1+\frac{dR}{cdt%
},\eqno
(4) 
$$
where the time derivative of the propagation range is evaluated at the
instant of wave emission.

The second-order Doppler effect can be given if the first-order formula of
the propagation range is used. By so doing, the Doppler frequency relation
becomes 
$$
\frac{f_{t}}{f_{r}}=1+\frac{dR_{t}}{cdt}+\frac{d}{c^{2}dt}({\bf v}_{e}\cdot 
{\bf R}_{t}).\eqno
(5) 
$$
Due to the relative motion between source and receiver, $d{\bf R}_{t}/dt=%
{\bf v}_{es}$ and hence $dR_{t}/dt=u_{es}$, where ${\bf v}_{es}$ ($={\bf v}%
_{e}-{\bf v}_{s}$) is the Newtonian relative velocity between receiver and
source at the instant of emission and $u_{es}$ ($=u_{e}-u_{s}$) is the
radial speed of the receiver with respect to the source. Thus, as given in
[4], the second-order Doppler frequency relation becomes 
$$
\frac{f_{t}}{f_{r}}=1+\frac{u_{es}}{c}+\frac{{\bf v}_{e}\cdot {\bf v}_{es}}{%
c^{2}}+\frac{{\bf a}_{e}\cdot {\bf R}_{t}}{c^{2}}.\eqno
(6) 
$$
It is noted that the transverse components of velocities are also involved
in the second-order Doppler shift. To the second order of normalized speed,
the inverse frequency ratio reads 
$$
\frac{f_{r}}{f_{t}}=1-\frac{u_{es}}{c}-\frac{{\bf v}_{e}\cdot {\bf v}_{es}}{%
c^{2}}-\frac{{\bf a}_{e}\cdot {\bf R}_{t}}{c^{2}}+\frac{u_{es}^{2}}{c^{2}}.%
\eqno
(7) 
$$
When the third-order Doppler effect needs to be considered, the second-order
formula of the propagation range should be used, as discussed later.

$\ $

\noindent {\large {\bf 3. Quantum Energy Variation and Resonant-Absorption
Condition}}

Based on the local-ether wave equation, it has been found that the mass of a
particle increases with its speed by the famous Lorentz mass-variation
factor, except that the speed is referred specifically to an ECI frame for
earthbound particles [3]. From this wave equation a first-order time
evolution equation is derived, which is similar to Schr\"{o}dinger's
equation except that the time derivative is referred specifically to the
local-ether frame and incorporates an extra multiplying term of the
mass-variation factor. Thereby, a quantum-mechanical approach has been
presented to show that the energies of quantum states of the matter wave
bounded in an atom or a molecule decrease with the inverse of the
speed-dependent mass. It is known that the frequency of light emitted from
or absorbed by an atom is equal to the transition frequency, which in turn
corresponds to the difference in energy between two involved quantum states.
Thus the state transition frequency $f$ of a moving atom decreases with its
speed by the speed-dependent mass-variation factor as [3] 
$$
f=f_{0}\sqrt{1-v^{2}/c^{2}},\eqno
(8) 
$$
where $v$ is the atom speed referred to the local-ether frame which is an
ECI frame for earthbound atoms, $f_{0}$ is the rest transition frequency of
an identical atom stationary in the local-ether frame, and the frequency $f$
is observed in the atom frame (with respect to which the atom is stationary)
such that no Doppler shift is involved.

The atomic clock rate depends on the transition frequency of the associated
atom and hence decreases with the atom speed by the mass-variation factor.
The preceding speed-dependent frequency-variation formula has been adopted
in [3] to account for the east-west directional anisotropy in the clock rate
demonstrated in the Hafele-Keating experiment with cesium atomic clocks
onboard an aircraft undergoing circumnavigation [5{]}. In this experiment
the observed clock rate is determined from the number of the atomic clock
ticks which in turn are detected and counted by a device which is comoving
with the clock without relative motion. Thus the clock rate is associated
with the quantum energy variation, but has nothing to do with the Doppler
frequency shift. Moreover, it has been used to account for the synchronism
and the clock-rate adjustment in GPS atomic clocks onboard earth's
satellites in circular orbits.

A speed-dependent frequency of the form of (8) (but of different physical
origin and reference frame of speed) was first introduced by Fitzgerald,
Lorentz, and Larmor before the advent of the special relativity [6] and was
later derived by assuming the length contraction [7] or the time dilation
[2]. In the local-ether model, the frequency-variation formula due to the
quantum effect is not expected to hold in all energy states in all atoms and
other forms of the speed-dependence are not precluded. This is in view of
the complication that the interactions which affect energy states are
versatile, such as electronic or nuclear, electric or magnetic, spin or
orbital, and intrinsic or external. In [3], the frequency-variation formula
is verified only for electronic states due to the electric scalar potential
in an atom or a molecule. Nevertheless, it is assumed that this formula
holds in the experiments examined in this investigation.

Then we proceed to consider the case where the source atom $s$ and the
receiver atom $e$ are located within the same local ether but are moving
respectively at velocities ${\bf v}_{s}$ and ${\bf v}_{e}$ with respect to
the local-ether frame. Due to the motions of the source and receiver, the
Doppler effect also causes a frequency shift. Suppose $f_{0s}$ and $f_{0e}$
are the rest transition frequencies of atoms $s$ and $e$, respectively. The
transmitted frequency $f_{t}$ of the wave emitted from the moving atom $s$
is related to the rest transition frequency $f_{0s}$ as 
$$
f_{t}=f_{0s}\sqrt{1-v_{s}^{2}/c^{2}}.\eqno
(9) 
$$
On the other hand, in order for the wave being resonantly absorbed by the
moving atom $e$, the rest transition frequency $f_{0e}$ should be related to
the received frequency $f_{r}$ as 
$$
f_{r}=f_{0e}\sqrt{1-v_{e}^{2}/c^{2}}.\eqno
(10) 
$$
The received frequency $f_{r}$ and the transmitted frequency $f_{t}$ in turn
are related to each other by the Doppler effect as shown in (6) or (7).

Thus, by taking both the Doppler frequency shift for electromagnetic wave
and the quantum energy variation of matter wave into account, we arrive at
the {\bf resonant-absorption condition} in terms of the rest transition
frequencies $f_{0s}$ and $f_{0e}$, which to the second order of normalized
speed is given by 
$$
f_{0e}\sqrt{1-v_{e}^{2}/c^{2}}=f_{0s}\sqrt{1-v_{s}^{2}/c^{2}}\left\{
1-u_{es}/c-({\bf v}_{e}\cdot {\bf v}_{es}+{\bf a}_{e}\cdot {\bf R}%
_{t}-u_{es}^{2})/c^{2}\right\} .\eqno
(11) 
$$
For the case where the receiver is comoving with the source (${\bf v}_{e}=%
{\bf v}_{s}$) without acceleration, the resonant-absorption condition
becomes trivially as $f_{0e}=f_{0s}$. This resonant-absorption condition
made its debut in [8].

When the resonant-absorption condition is not met, the off-resonance
absorption tends to be much weaker. The deviation in frequency from the
resonant-absorption condition is given by the difference between the
received frequency and the transition frequencies of the absorbing atoms in
motion. That is, 
$$
\delta f=f_{r}-f_{0e}\sqrt{1-v_{e}^{2}/c^{2}}.\eqno
(12) 
$$
By using (9), the frequency deviation can be written as 
$$
\delta f=\triangle f_{D}-\triangle f_{Q},\eqno
(13) 
$$
where $\triangle f_{D}$ and $\triangle f_{Q}$ denote the frequency shifts
due to the Doppler and the quantum effects and are given by 
$$
\triangle f_{D}=f_{r}-f_{t}\eqno
(14) 
$$
and 
$$
\triangle f_{Q}=f_{0e}\sqrt{1-v_{e}^{2}/c^{2}}-f_{0s}\sqrt{1-v_{s}^{2}/c^{2}}%
,\eqno
(15) 
$$
respectively. In words, $\triangle f_{Q}$ is the difference in the
speed-dependent transition frequency between the receiver and source atoms
in motion. From (4) it is seen that the fractional Doppler frequency shift
is given by 
$$
\frac{\triangle f_{D}}{f_{r}}=-\frac{dR}{cdt}.\eqno
(16) 
$$
The resonant-absorption condition corresponds to the deviation $\delta f=0$,
which means that the frequency shift due to the Doppler effect for
electromagnetic wave propagating in free space exactly cancels the one due
to the quantum effect of matter wave bounded in atom.

$\ $

\noindent {\large {\bf 4. Reexamination of Resonant-Absorption Experiments}}

Based on the local-ether model, we reexamine the resonant-absorption
experiments reported in the literature. Thereby, we present
reinterpretations of the Ives-Stilwell experiment,{\rm \ }the output
frequency from ammonia masers, and of the M\"{o}ssbauer rotor experiment and
explore the associated spatial isotropy, Galilean relativity, and their
breakdowns. The first experiment is associated with a radial relative motion
($u_{es}\simeq \pm v_{es}$) and the last two are with a transverse relative
motion ($u_{es}=0$).

$\ $

\noindent {\bf 4.1. Ives-Stilwell experiment}

Consider the case where the source is moving at a velocity ${\bf v}_{s}$ and
the receiver is stationary, both with respect to the local-ether frame. Thus 
$a_{e}=0$, $v_{e}=0$, and $u_{es}=-u_{s}$. Then the second-order
resonant-absorption condition becomes a simpler form of 
$$
f_{0e}=f_{0s}\frac{\sqrt{1-v_{s}^{2}/c^{2}}}{1-u_{s}/c}.\eqno
(17) 
$$
For this case $f_{r}=f_{0e}$ and thus this formula is identical to that
derived in different ways based on the transformation of four-vectors [9] or
on the time dilation [2], except the difference in reference frame of
velocity ${\bf v}_{s}$. According to the local-ether model, the term $%
(1-u_{s}/c)$ in the preceding formula is due to the Doppler frequency shift
to the second order, while the term $\sqrt{1-v_{s}^{2}/c^{2}}$ is due to the
quantum energy variation.

A similar frequency shift has been demonstrated in the Ives-Stilwell
experiment, where light is emitted from fast-moving hydrogen atoms in an
excited state and is absorbed by a spectrograph used to measure the received
frequency [1, 10]. The radiation from the moving atoms is reflected from two
mirrors and then the two reflected light beams are guided to a photographic
plate where the frequencies are recorded for measurement and comparison. The
two mirrors are arranged in such a way that the atoms are moving toward one
of them and away from the other [10]. Due to the Doppler effect with the
relative motions between the emitting atoms and the mirrors, the light beams
reflected from the mirrors tend to shift in frequency. However, by virtue of
no relative motions among the mirrors, the spectrograph, and the other
components of experimental setup, no further Doppler effect is introduced
after the reflection.

Note that even for a geostationary experimental setup, the receiver is not
stationary with respect to the associated local-ether frame due to earth's
rotation. Consider the laboratory frame in which the setup and the receiver
are stationary. Suppose that the emitting atoms are moving at a velocity $%
{\bf v}_{t}$ with respect to the laboratory frame, which in turn moves at a
velocity ${\bf v}_{0}$ with respect to an ECI frame. Thus ${\bf v}_{e}={\bf v%
}_{0}$, ${\bf v}_{s}={\bf v}_{t}+{\bf v}_{0}$, and ${\bf v}_{es}=-{\bf v}%
_{t} $. In the experiment the atoms move nearly in the radial direction and
the directions of the two propagation paths are antiparallel. Thus $%
u_{es}=\mp v_{t}\cos \theta $, where $\theta $ denotes the small angle of $%
{\bf v}_{t}$ from $\hat{R}_{t}$, $\pm \hat{R}_{t}$ represents the direction
from the emitting atoms to the reflecting point on either mirror, and the
upper or the lower sign applies for the light beam reflected from the mirror
which the atoms approach or recede from, respectively. Then the
resonant-absorption condition (11) leads to that the received frequencies
for the two reflected light beams are 
$$
f_{r\pm }=f_{0s}\sqrt{1-({\bf v}_{t}+{\bf v}_{0})^{2}/c^{2}}\left\{ 1\pm
v_{t}\cos \theta /c+({\bf v}_{0}\cdot {\bf v}_{t}+v_{t}^{2}\cos ^{2}\theta
)/c^{2}\right\} .\eqno
(18) 
$$
It is noted that the first-order term is independent of the laboratory
velocity ${\bf v}_{0}$. If the velocity ${\bf v}_{0}$ is neglected, the
preceding formula reduces to (17).

To the second order of normalized speed, the received frequencies become 
$$
f_{r\pm }=f_{0s}\left\{ 1\mp v_{t}\cos \theta
/c+(v_{t}^{2}+v_{0}^{2})/2c^{2}\right\} ^{-1}.\eqno
(19) 
$$
Both the Doppler and the quantum effects contribute to the second-order
terms. It is noted that the dot-product term ${\bf v}_{0}{\bf \cdot v}_{t}$
appears both in the Doppler and the quantum effects and then cancels out. In
the Ives-Stilwell experiment the speed $v_{t}$ of hydrogen atoms is of the
order of 10$^{6}$ m/sec, which is achieved by electrostatically accelerating
hydrogen ions with a voltage of tens of kV. Thus the atom speed $v_{t}$ is
much higher than the laboratory speed $v_{0}$ which is due to earth's
rotation and, perhaps, to the motion of a vehicle carrying the experimental
setup. Thus the preceding frequency formula can be substantially independent
of the laboratory velocity ${\bf v}_{0}$.

The first- and the second-order fractional frequency shifts are associated
with the difference between and with the sum of the two received
frequencies, respectively. Thus, in terms of wavelengths, the normalized
speed and its square can be given by 
$$
\frac{v_{t}}{c}=\frac{1}{2\cos \theta }\frac{\lambda _{R}-\lambda _{B}}{%
\lambda _{0}}\eqno
(20) 
$$
and 
$$
\frac{v_{t}^{2}}{c^{2}}=\left( \frac{\lambda _{R}+\lambda _{B}}{\lambda _{0}}%
-2\ -\frac{v_{0}^{2}}{c^{2}}\right) ,\eqno
(21) 
$$
where $\lambda _{B}=c/f_{r+}$, $\lambda _{R}=c/f_{r-}$, and $\lambda
_{0}=c/f_{0s}$. By measuring $\lambda _{R}$ and $\lambda _{B}$, the first-
and the second-order fractional shifts can be calculated and compared, with
a knowledge of $\theta $, $\lambda _{0}$, and $v_{0}$. This provides a
crucial means to test the minor second-order frequency shift. It can be
expected that the deviation of the sum $\lambda _{R}+\lambda _{B}$ from $%
2\lambda _{0}$ is quite small. By using a high-precision spectrograph, it
has been measured that $\lambda _{R}=6618.808$ (in angstrom) and $\lambda
_{B}=6507.253$ [10]. Meanwhile, it is known that $\lambda _{0}=6562.793$ and 
$\theta =0.075$ (in rad). Then formulas (20) and (21) lead to $%
v_{t}^{2}/c^{2}=7.26\times 10^{-5}$ and 7.24$\times 10^{-5}$, respectively,
where the term $v_{0}^{2}/c^{2}$ is as small as 10$^{-12}$ and hence is
omitted. The agreement between the two data is quite good, since the
discrepancy is as small as the round-off in the calculation. According to
the local-ether model, the wavelength $\lambda _{0}$ corresponds to a
transition frequency of a hydrogen atom stationary with respect to an ECI
frame. This seems to be in accord with the description that $\lambda _{0}$
is the wavelength of the H$_{\alpha }$ line of the Balmer series observed
in\ a reference frame at rest with respect to the radiating atom [10].
However, the value of $\lambda _{0}$ could be actually measured from
geostationary atoms in a geostationary laboratory. Even so, the fractional
discrepancy in $\lambda _{0}$ due to the difference in reference frame is
also of the order of $v_{0}^{2}/c^{2}$. Thus the local-ether model is in
accord with the self-consistency between the first- and the second-order
frequency shifts demonstrated in the Ives-Stilwell experiment.

$\ $

\noindent {\bf 4.2. Output frequency from ammonia masers}

Next, we consider the case where the source is moving transverse to the
propagation path. In the ammonia maser, the molecular beam is injected into
a resonant cavity and emits microwave due to quantum state transition [11,
2]. The output of the microwave comes through a waveguide coupled to the
cavity via a small aperture. Suppose that the cavity and the receiver are
stationary in the laboratory frame of velocity ${\bf v}_{0}$ and that the
emitting ammonia molecular beam moves at a velocity ${\bf v}_{t}$ with
respect to the cavity. Moreover, suppose that the output waveguide is thin
and long and is positioned with its longitudinal axis being perpendicular to
the direction of molecular beam represented by ${\bf v}_{t}$. Thus the
direction of the propagation path of the microwave is virtually parallel to
the longitudinal axis of the output waveguide and hence is transverse to the
molecule velocity, that is, ${\bf v}_{t}{\bf \cdot }\hat{R}_{t}=0$.

Thus ${\bf v}_{e}={\bf v}_{0}$, ${\bf a}_{e}=0$, ${\bf v}_{s}={\bf v}_{t}+%
{\bf v}_{0}$, ${\bf v}_{es}=-{\bf v}_{t}$, and $u_{es}=0$. Then the
resonant-absorption condition (11) leads to the received frequency 
$$
f_{r}=f_{0s}\sqrt{1-({\bf v}_{t}+{\bf v}_{0})^{2}/c^{2}}\left\{ 1+{\bf v}%
_{0}\cdot {\bf v}_{t}/c^{2}\right\} .\eqno
(22) 
$$
The Doppler effect is presented by the dot-product term ${\bf v}_{0}{\bf %
\cdot v}_{t}$. Thus, by taking both the Doppler and the quantum effects into
account, the resonant-absorption condition leads to that the received
frequency at the output waveguide of the maser is given by 
$$
\frac{f_{r}}{f_{0s}}=1-\frac{v_{t}^{2}+v_{0}^{2}}{2c^{2}}.\eqno
(23) 
$$
It is noted again that the dot-product term ${\bf v}_{0}{\bf \cdot v}_{t}$
appears both in the Doppler and the quantum effects and then cancels out.
Consequently, the output frequency from the maser is independent of the
directions of velocities ${\bf v}_{t}$ and ${\bf v}_{0}$. Moreover, the
resonant-absorption condition leads to 
$$
\frac{f_{0e}}{f_{0s}}=1-\frac{v_{t}^{2}}{2c^{2}}.\eqno
(24) 
$$
It is seen that this frequency relation is even independent of the
laboratory velocity ${\bf v}_{0}$.

It is noticed that a term quite similar to the second-order Doppler effect
in (22) has been derived alternatively from a classical approach, aside from
the reference frame of the laboratory velocity [11, 2]. Further, a similar
cancellation of the terms associated with ${\bf v}_{0}$ has been accounted
for by an alternative approach, where a speed-dependent frequency based on
the time dilation is used [12]. Anyway, it has been demonstrated
experimentally that the beat frequency between two masers with molecular
beams moving in opposite directions is substantially zero and is almost
independent of earth's motions [11]. Thus the local-ether model is in accord
with the experiments with ammonia masers.

$\ $

\noindent {\bf 4.3. M\"{o}ssbauer rotor experiment}

Then we consider the case where both the source and the absorber are moving
in the laboratory frame. The M\"{o}ssbauer rotor experiment is based on the
recoilless gamma-ray nuclear resonance absorption known as the M\"{o}ssbauer
effect [13-17]. The source and the absorber of gamma ray are placed
separately on a rotating rod. Both the quantum energy variation of nuclear
states and the Doppler frequency shift of wave propagation cause the
frequency shift. A frequency deviation from the resonant-absorption
condition exhibits itself with an increase in the scattering of gamma ray,
which in turn can be measured by counters fixed at the laboratory or at the
rod.

Suppose that the absorber and the source atoms in the M\"{o}ssbauer rotor
experiment rotate about an axis with linear velocities ${\bf v}_{r}$ and $%
{\bf v}_{t}$, respectively, where the axis is stationary in a laboratory
frame which in turn moves at a velocity ${\bf v}_{0}$ with respect to the
local-ether frame. In the rotor experiment, the directed path length ${\bf R}%
_{t}$ from the source to the absorber is always perpendicular both to ${\bf v%
}_{r}$ and ${\bf v}_{t}$. Moreover, the path length $R_{t}$ is fixed, while
the direction $\hat{R}_{t}$ is changing with time. Thus $dR_{t}/dt=0$ and $d%
{\bf R}_{t}/dt={\bf v}_{rt}$, where ${\bf v}_{rt}={\bf v}_{r}-{\bf v}_{t}$.

To the second order of normalized speed, the fractional Doppler frequency
shift (16) due to the motions of the source and receiver becomes 
$$
\frac{\triangle f_{D}}{f_{r}}=-\frac{d}{cdt}\left( R_{t}+\frac{1}{c}{\bf v}%
_{0}\cdot {\bf R}_{t}\right) =-\frac{{\bf v}_{0}\cdot {\bf v}_{rt}}{c^{2}},%
\eqno
(25) 
$$
where the orthogonality ${\bf v}_{r}{\bf \cdot R}_{t}=0$ has been made use
of and the derivative $d{\bf v}_{0}/dt$ is omitted. It is seen that this
frequency shift depends on the laboratory velocity ${\bf v}_{0}$. It is
noted that due to the path length $R_{t}$ being a constant, the first-order
term of normalized speed vanishes in the preceding formula. Thus, to the
third order of normalized speed, the fractional shift $\triangle
f_{D}/f_{0e} $ is identical to the preceding formula of $\triangle
f_{D}/f_{r}$.

Consider the case where the source and the absorber have the identical rest
transition frequency $f_{0}$. Thus the fractional frequency shift due to the
quantum effect is 
$$
\frac{\triangle f_{Q}}{f_{0}}=\sqrt{1-({\bf v}_{r}+{\bf v}_{0})^{2}/c^{2}}-%
\sqrt{1-({\bf v}_{t}+{\bf v}_{0})^{2}/c^{2}}\simeq \frac{1}{2c^{2}}(-2{\bf v}%
_{rt}\cdot {\bf v}_{0}-v_{r}^{2}+v_{t}^{2}).\eqno
(26) 
$$
It is noted that the dot-product term ${\bf v}_{rt}{\bf \cdot v}_{0}$
appears again. Consequently, by taking both the Doppler and the quantum
effects into account to the second order of normalized speed, the fractional
frequency deviation from the resonant absorption is 
$$
\frac{\delta f}{f_{0}}=\frac{v_{r}^{2}-v_{t}^{2}}{2c^{2}}.\eqno
(27) 
$$
It is noted that the dot-product term ${\bf v}_{0}\cdot {\bf v}_{rt}$
cancels out and hence the frequency deviation is independent of the
directions of velocities ${\bf v}_{r}$ and ${\bf v}_{t}$. Furthermore, this
frequency deviation is independent of the laboratory velocity ${\bf v}_{0}$.

It is noticed again that a term quite similar to the second-order Doppler
effect in (25) has been derived alternatively from a classical approach,
aside from the reference frame of the laboratory velocity [14, 16, 17].
Further, a similar cancellation of the terms with ${\bf v}_{0}$ has been
accounted for by an alternative approach, where a speed-dependent frequency
based on the length contraction is used [16]. Anyway, this fractional
frequency deviation has been demonstrated in experiments with various
rotation rates and various positions of the source and the absorber at the
rod [13-15]. Moreover, it has been found that between the two geostationary
counters oriented to detect the gamma rays propagating south- and northward,
respectively, there is no substantial difference in the measured frequency
deviation [14]. Thus the frequency deviation can be independent of the
orientation of counter. Based on these, the local-ether model is also in
accord with the M\"{o}ssbauer rotor experiment.

$\ $

\noindent {\bf 4.4. Spatial isotropy, Galilean relativity, and their
breakdowns}

According to the local-ether model, the speed of an earthbound atom is
referred to an ECI frame. Thus the quantum energy in an atom is entirely
independent of earth's orbital motion around the Sun or whatever. Further,
the quantum energy can even be independent of earth's rotation, if the atom
speed remains a constant during the rotation. Such a constant-speed
condition is met by a geostationary atom, by an atom moving at a fixed
velocity with respect to ground at a substantially fixed latitude, or by an
atom moving in a circular satellite orbit around the Earth. For an atom
satisfying the constant-speed condition, the energies of quantum states and
hence the transition frequency between two states are independent of the
orientation and position of the Earth in space, whatever the dependence of
quantum energy on the speed. This spatial isotropy of transition frequency
with respect to earth's motions has been adopted in [3] to account for the
hourly and daily frequency stability in geostationary atoms in the
Hughes-Drever experiment [18] and for the high synchronism among the various
GPS atomic clocks moving in circular orbits [19]. On the other hand, for two
atomic clocks moving even at an identical speed but in different directions
with respect to the ground, their speeds with respect to an ECI frame tend
to be different. Thereby, this isotropy breaks down, as demonstrated in the
east-west directional anisotropy in atomic clock rate in the Hafele-Keating
experiment [5{]}.

The applicability of the aforementioned spatial isotropy can be extended to
the cases where electromagnetic wave as well as matter wave is involved.
According to the local-ether model, the results of an earthbound experiment
are entirely independent of earth's orbital motion. Further, the results can
be independent of earth's rotation, so long as the atom speeds $v_{e}$, $%
v_{s}$, $u_{e}$, and $u_{s}$ along with the terms ${\bf v}_{e}{\bf \cdot v}%
_{s}$ and ${\bf a}_{e}{\bf \cdot R}_{t}$ are invariant under earth's
rotation. Thereby, the frequency-variation formula (8), the
propagation-range formula (2), and the resonant-absorption condition (11)
remain unchanged under earth's rotation and hence the associated phenomena
exhibit the spatial isotropy. In the Ives-Stilwell experiment and the
ammonia maser, it is seen that the speeds $v_{e}$ and $v_{s}$ are invariant
under earth's rotation and the acceleration ${\bf a}_{e}$ is zero. Moreover,
the terms $u_{e}$, $u_{s}$, and ${\bf v}_{e}{\bf \cdot v}_{s}$ are invariant
under earth's rotation, since ${\bf v}_{e}$, ${\bf v}_{s}$, and ${\bf R}_{t}$
all change in a coordinated way with earth's rotation. Thus the spatial
isotropy can be expected in these experiments and has been demonstrated in
the hourly and daily stability of the output frequency from ammonia masers
[11].

The spatial isotropy with respect to earth's motions can be generalized with
a step forward. As the velocities ${\bf v}_{e}$ and ${\bf v}_{s}$ are
written in the laboratory frame of velocity ${\bf v}_{0}$, the terms ${\bf v}%
_{r}{\bf \cdot v}_{0}$, ${\bf v}_{t}{\bf \cdot v}_{0}$, and $v_{0}^{2}$
emerge in the relevant formulas. It has been indicated that the dot-product
term ${\bf v}_{r}{\bf \cdot v}_{0}$ or ${\bf v}_{t}{\bf \cdot v}_{0}$ can be
common in the Doppler and the quantum effects and then cancels out.
Consequently, the results of the experiment are independent of the
orientation of the setup with respect to the ground. Thus the experiment
possesses a spatial isotropy with respect to the setup orientation as well
as to earth's motions. This kind of spatial isotropy is observed in the
three experiments examined so far. Thereby, the output frequency from the
ammonia maser is independent of the direction of the molecule velocity and
hence no observable beat frequency between masers with different
orientations can be expected, as demonstrated experimentally in [11].
Moreover, the frequency deviation from the resonant absorption in the
M\"{o}ssbauer rotor experiment can be expected to be stable under earth's
rotation and be identical for counters with different orientations, as
demonstrated experimentally in [14].

Further, the squared term $v_{0}^{2}/c^{2}$ in some frequency formulas
cancels out or is too small to detect. As a consequence, the experimental
results become independent of the laboratory velocity ${\bf v}_{0}$ and
hence comply with Galilean relativity, in spite of the restriction on the
reference frame of the particle and the propagation velocities. Thus, when
the whole experimental setup is put on a vehicle moving smoothly with
respect to the ground, the measurement results can be independent of the
ground velocity of the vehicle, as well as of the setup orientation and
earth's motions. Thereby, the spatial isotropy is generalized to the
compliance with Galilean relativity. In this way, it is seen that the
M\"{o}ssbauer rotor experiment complies with Galilean relativity. Moreover,
Galilean relativity can also be observed in the Ives-Stilwell experiment and
in the output frequency from the ammonia maser, as the minute squared term
is difficult to detect. Anyway, these experiments preserve the spatial
isotropy with respect to the setup orientation and earth's motions.

However, when the measurement precision is improved such that higher-order
effects can be detected, some minute terms may emerge to make Galilean
relativity or even the spatial isotropy break down. The just-mentioned
squared term $v_{0}^{2}/c^{2}$ in the Ives-Stilwell experiment and the
ammonia maser is an example. For another, we reexamine the M\"{o}ssbauer
rotor experiment to a higher order. In order to derive the Doppler effect to
the third order of normalized speed, the second-order propagation-range
formula (2) is needed, which for the M\"{o}ssbauer rotor experiment becomes 
$$
R(t)=R_{t}\left\{ 1+\frac{1}{c}u_{0}+\frac{1}{2c^{2}}[%
u_{0}^{2}+(v_{r}^{2}+v_{0}^{2}+2{\bf v}_{r}\cdot {\bf v}_{0})+{\bf a}%
_{e}\cdot {\bf R}_{t}]\right\} .\eqno
(28) 
$$
For a rotation at a fixed angular velocity, the terms of $v_{r}$ and ${\bf a}%
_{e}{\bf \cdot R}_{t}$ as well as $v_{0}$ are constants during rotation.
Thus the time rate of change of the propagation range $R$ is given by 
$$
\frac{dR(t)}{dt}=R_{t}\frac{d}{dt}\left\{ \frac{1}{c}u_{0}+\frac{1}{2c^{2}}[%
u_{0}^{2}+2{\bf v}_{r}\cdot {\bf v}_{0}]\right\} .\eqno
(29) 
$$
Then, to the third order of normalized speed, the fractional Doppler
frequency shift due to wave propagation can be given from (16) as 
$$
\frac{\triangle f_{D}}{f_{0}}=-\frac{dR}{cdt}=-\frac{1}{c^{2}}{\bf v}%
_{rt}\cdot {\bf v}_{0}-\frac{1}{c^{3}}[({\bf v}_{rt}\cdot {\bf v}%
_{0})u_{0}-v_{r}v_{rt}u_{0}],\eqno
(30) 
$$
where we have made use of $d{\bf v}_{r}/dt=-\hat{R}_{t}v_{r}v_{rt}/R_{t}$
and of $\triangle f_{D}/f_{0}=\triangle f_{D}/f_{r}$ to the third order. It
is seen that $\triangle f_{D}=0$ when $v_{0}=0$.

By taking both the Doppler and the quantum effects into account to the third
order of normalized speed, the fractional frequency deviation from the
resonant absorption is 
$$
\frac{\delta f}{f_{0}}=\frac{1}{2c^{2}}(v_{r}^{2}-v_{t}^{2})-\frac{1}{c^{3}}(%
{\bf v}_{rt}\cdot {\bf v}_{0}-v_{r}v_{rt})u_{0}.\eqno
(31) 
$$
It is noted that the dot-product term ${\bf v}_{rt}{\bf \cdot v}_{0}$
survives in the third-order effect, although it cancels out in the
second-order one. The presence of this term then makes the spatial isotropy
with respect to the setup orientation and hence Galilean relativity break
down, although it is quite small in magnitude. Accordingly, the frequency
deviations tend to be different as measured in counters with different
orientations with respect to the ground, when they can be measured to the
third order. Thereby, a directional anisotropy in the frequency deviation is
predicted. On the other hand, the spatial isotropy with respect to earth's
rotational and orbital motions is preserved.

$\ $

\noindent {\large {\bf 5. Conclusion}}

Based on the local-ether wave equation for matter wave, the energies of
quantum states and hence the state transition frequency in an atom decrease
with the atom speed by the mass-variation factor, where the speed is
referred to the associated local-ether frame. Further, for the situation
where the source and the receiver atoms move at different velocities, the
local-ether resonant-absorption condition is presented by taking both the
Doppler frequency shift for electromagnetic wave and the quantum energy
variation of matter wave into account. It is shown that the
resonant-absorption condition accounts for the Ives-Stilwell experiment, the
output frequency from ammonia masers, and the M\"{o}ssbauer rotor experiment
in a consistent way.

Based on the local-ether model, it is evident that the phenomena due both to
electromagnetic and matter waves do not at all depend on earth's orbital
motion around the Sun or others. This immediately accounts for the null
effect of earth's orbital motion in various earthbound phenomena. Further,
for a geostationary experimental setup, the results can be independent of
earth's rotation. This accounts for the spatial isotropy found in the
stability of frequency{\rm \ }in the Hughes-Drever experiment and the
ammonia maser. The spatial isotropy also holds for the atomic clocks onboard
earth's satellites moving in circular orbits and hence accounts for the high
synchronism among the various GPS atomic clocks. Further, in the formula of
the Doppler or the quantum effect, it is seen that the major term associated
with the laboratory velocity is a dot product between this velocity and that
of the emitting or absorbing atom. This term appears both in the Doppler
frequency shift and the transition frequency variation and then cancels out.
Thus the experimental results become invariant with respect to the setup
orientation, as well as to earth's motions. This spatial isotropy has been
demonstrated in the ammonia maser where the output frequency is independent
of the direction of the molecule velocity and in the M\"{o}ssbauer rotor
experiment where the frequency deviation from the resonant absorption is
independent of the orientation of the counter. Further, the frequency
deviation is independent of the laboratory velocity and hence the
M\"{o}ssbauer rotor experiment complies with Galilean relativity.

However, the aforementioned dot-product term with the laboratory velocity
may survive in some formulas. Thus the spatial isotropy with respect to the
setup orientation and hence Galilean relativity break down. This breakdown
is in accord with the east-west directional anisotropy in atomic clock rate
demonstrated in the Hafele-Keating experiment. By examining the
resonant-absorption condition in the M\"{o}ssbauer rotor experiment to the
third order, such a breakdown is also found. Thereby, it is expected that
the frequency deviation tends to be different as measured in counters with
different orientations. This predicted directional anisotropy in the
frequency deviation may provide a mean to test the local-ether model, if the
measurement precision can be raised to the third order of normalized speed.

$\ $

$\ $

\noindent {\large {\bf References}}

\begin{itemize}
\item[{\lbrack 1]}]  {H.E. Ives and G.R. Stilwell, {\it J. Opt. Soc. Am.} 
{\bf 28}, 215 (1938); {\it J. Opt. Soc. Am.} {\bf 31}, 369 (1941).}

\item[{\lbrack 2]}]  A.P. French{, {\it Special Relativity} (Chapman \&
Hall, New York, 1968), ch. 5.}

\item[{\lbrack 3]}]  C.C. Su, {\it Eur. Phys. J. B} {\bf 24}, 231 (2001).

\item[{\lbrack 4]}]  C.C. Su, {\it Eur. Phys. J. C} {\bf 21}, 701 (2001); 
{\it Europhys. Lett}. {\bf 56}, 170 (2001).

\item[{\lbrack 5]}]  {J.C. Hafele and R.E. Keating, {\it Science} {\bf 177},
166 (1972); {\bf 177}, 168 (1972).}

\item[{\lbrack 6]}]  {H.E. Ives, {\it J. Opt. Soc. Am}. {\bf 27}, 263 (1937).%
}

\item[{\lbrack 7]}]  R.J. Kennedy and E.M. Thorndike, {\it Phys. Rev}. {\bf %
42}, 400 (1932).

\item[{\lbrack 8]}]  C.C. Su, in{\ {\it Bull. Am. Phys. Soc.}} (Apr. 2001)%
{\it ,} p. 99{.}

\item[{\lbrack 9]}]  {L.D. Landau and E.M. Lifshitz, {\it The Classical
Theory of Fields}} {(Pergamon, New York, 1975), sect. 48.}

\item[{\lbrack 10]}]  {H.I. Mandelberg and L. Witten, {\it J. Opt. Soc. Am.} 
{\bf 52}, 529 (1962).}

\item[{\lbrack 11]}]  {J.P. Cedarholm, G.F. Bland, B.L. Havens, and C.H.
Townes, {\it Phys. Rev. Lett}}. {\bf 1}, 342 (1958).

\item[{\lbrack 12]}]  {J.P. Cedarholm and C.H. Townes, {\it Nature} {\bf 181}%
, 1350 (1959).}

\item[{\lbrack 13]}]  {H.J. Hay, J.P. Schiffer, T.E. Cranshaw, and P.A.
Egelstaff, }{\it Phys. Rev. Lett}. {\bf 4}, 165 (1960).

\item[{\lbrack 14]}]  {D.C. Champeney, G.R. Isaak, and A.M. Khan, {\it Phys.
Lett}. {\bf 7}, 241 (1963).}

\item[{\lbrack 15]}]  {K.C. Turner and H.A. Hill, {\it Phys. Rev}}. {\bf 134}%
, B252 (1964).

\item[{\lbrack 16]}]  {M. Ruderfer, }{\it Phys. Rev. Lett}. {\bf 5}, 191
(1960); {\bf 7}, 361 (1961).

\item[{\lbrack 17]}]  {J.D. Jackson, {\it Classical Electrodynamics} (Wiley,
New York, 1975), ch. 11.}

\item[{\lbrack 18]}]  {M.P. Haugan and C.M. Will, {\it Phys. Today} {\bf 40}%
, 69 (1987).}

\item[{\lbrack 19]}]  T. Logsdon, {\it The NAVSTAR Global Positioning System}
(Van Nostrand Reinhold, New York, 1992), chs. 2 and 11.
\end{itemize}

\end{document}